\documentclass[journal]{IEEEtran}

\IEEEoverridecommandlockouts
\usepackage{amsmath}
\usepackage{amsfonts}
\usepackage{amssymb}
\usepackage{graphicx}
\usepackage{epstopdf}
\usepackage{tikz}
\usetikzlibrary{arrows.meta,positioning}
\usetikzlibrary{backgrounds}
\usepackage{pgfplots}
\pgfplotsset{compat=newest}
\usepackage{pgfplotstable}
\usepgfplotslibrary{groupplots}
\usepackage{filecontents}
\usepackage{cite}
\usepackage{algorithm}
\usepackage{algpseudocode}
\usepackage{blindtext}
\usepackage{enumitem}
\usepackage{pdfpages}
\usepackage{bbm}
\usepackage{bibentry}
\usepackage{amsthm}
\usepackage{color}
\usepackage{mathtools}
\usepackage[table,xcdraw]{xcolor}
\usepackage{multirow}

\usepackage{comment}

\makeatletter
\newcommand\fs@betterruled{%
  \def\@fs@cfont{\bfseries}\let\@fs@capt\floatc@ruled
  \def\@fs@pre{\vspace*{8pt}\hrule height.8pt depth0pt \kern2pt}%
  \def\@fs@post{\kern2pt\hrule\relax}%
  \def\@fs@mid{\kern2pt\hrule\kern2pt}%
  \let\@fs@iftopcapt\iftrue}
\floatstyle{betterruled}
\restylefloat{algorithm}
\makeatother

\definecolor{DarkGreen}{RGB}{0,150,0}

\usepackage{balance}

\begin{document}
\title{Robust and Resilient Networks with Integrated Sensing, Communication and Computation}

% \mc Overview and Design Aspects for 

\author{Ming-Chun Lee,~\IEEEmembership{Member,~IEEE}, 
Christian Eckrich,~\IEEEmembership{Student Member,~IEEE},
Vahid Jamali,~\IEEEmembership{Senior Member,~IEEE},\newline 
Yu-Chih Huang,~\IEEEmembership{Senior Member,~IEEE}, 
Arash Asadi,~\IEEEmembership{Senior Member,~IEEE}, 
Li-Chun Wang,~\IEEEmembership{Fellow,~IEEE} 
% Institute of Communications Engineering\\
% National Yang Ming Chiao Tung University\\
% Email: mingchunlee@nycu.edu.tw
\thanks{Lee, Huang, and Wang's work was supported in part by the National Science and Technology Council (NSTC) of Taiwan under grants 114-2628-E-A49-011-MY3 and 114-2224-E-A49-002-. Eckrich and Jamali’s work was supported in part by the Deutsche Forschungsgemeinschaft (DFG, German Research Foundation) under project number JA 3104/4-1, in part by the LOEWE Initiative, Hesse, Germany, within the emergenCITY Center [LOEWE/1/12/519/03/05.001(0016)/72], and in part by German Federal Ministry of Research, Technology and Space (BMFTR) within the Project ‘‘Open6GHub’’ under Grant 16KISK014. (M.-C. Lee and C. Eckrich are co-first authors.) (Corresponding author: M.-C. Lee)}
\thanks{M.-C. Lee, Y.-C Huang, and L.-C. Wang are with Institute of Communications Engineering, National Yang Ming Chiao Tung University, Hsinchu 30010, Taiwan. (email: mingchunlee@nycu.edu.tw; jerryhuang@nycu.edu.tw; wang@nycu.edu.tw)
}
\thanks{C. Eckrich and V. Jamali are with Department of Electrical Engineering and Information
Technology, Technical University of Darmstadt, 64283 Darmstadt, Germany (e-mail: christian.eckrich@tu-darmstadt.de; vahid.jamali@tu-darmstadt.de).
}
\thanks{A. Arash is with the Embedded Systems Group, TU Delft, 2628 CD Delft, The Netherlands (email: a.asadi@tudelft.nl).}
}

\maketitle
\begin{abstract}
%\boldmath
Emerging applications such as networked robotics, intelligent transportation, smart factories, and virtual and augmented reality demand integrated perception and connectivity enabled by wireless communication. This has driven growing interests in integrated sensing, communication, and computation (ISCC) systems, with a primary focus on their efficient co-designs. However, as ISCC systems increasingly support critical applications, they must not only deliver high performance but also demonstrate robustness and resilience. In this context, robustness refers to a system's ability to maintain performance under uncertainties, while resilience denotes its capacity to sustain a minimum level of service in the face of major disruptions. To address this gap, this article presents an overview of ISCC systems from the perspectives of robustness and resilience under limited resources. First, key concepts related to these properties are introduced in the ISCC context. Subsequently, design approaches for realizing robust and resilient ISCC networks are discussed. Finally, the article concludes with the discussions of a case study and open research problems in this area.
\end{abstract}

\IEEEpeerreviewmaketitle
% \begin{IEEEkeywords}
% Edge-caching and edge-computing, integration of caching, computing and communication, robust optimization, uncertainty.
% \end{IEEEkeywords}

\section{Introduction}

Sensing and communications through wireless devices and networks have evolved largely independently over the past decades, while each playing a vital role in enabling new wireless services. Recently, the emergence of advanced applications in the sixth-generation (6G) wireless networks, such as intelligent healthcare, Industry 5.0, autonomous transportation, and smart cities, has created a pressing need for these two functions to work jointly.
This has motivated intense studies on integrated sensing and communication (ISAC) in past years \cite{Lu2024ISAC_Open}. The relevant standardization progress for ISAC has been launched by 3GPP within the scope of IMT-2030 \cite{3GPP_Rel19_ISAC,IMT_2030}. 

Yet, the demands of these next-generation applications extend beyond the joint operation of sensing and communication. In the age of artificial intelligence (AI), networks must not only perceive and connect, but also reason and act intelligently.
This then calls for edge-AI, which brings the computational capability closer to where data is generated \cite{Wen2024_ISACAI}. 
However, the effectiveness of AI-based decision-making highly relies on the timely acquisition of environmental information and their delivery to the AI models. Therefore, the true intelligence comes from the integration of sensing, communication, and computing (ISCC), which allows applications to efficiently utilize the sensing information along with communication and computation resources to make smart decisions and reactions within a short time \cite{Wen2024_ISCC}.

While 6G delivers more intelligent and pervasive ISCC services to the society, the underlying network architecture is becoming increasingly complex to support this interconnected set of functionalities.
This complexity can at the same time lead to a more sensitive and vulnerable network, where disruptions and failures in any of the sensing, communication, and computing components could lead to service interruptions and, consequently, severe societal impacts
\cite{Wen2024_ISCC}. In this context, the robustness and resilience aspects of the network have started to 
draw attention
\cite{KhaloopourResilience6G2024,reifert2024resilience}. ``Robustness'' refers to the ability of the system to tolerate a certain level of uncertainty without experiencing substantial performance degradation, that is the system should maintain its functionality within a pre-defined range of uncertainty. In contrast, ``resilience'' refers to the capability of systems to maintain a minimum performance even facing major disturbances, such as technical failures, human errors, jamming attacks or other unexpected strikes/changes, as well as to recover from these disruptions to an acceptable operational state over time \cite{KhaloopourResilience6G2024}.

While robustness and resilience in wireless networks have gained attention, they remain underexplored in the context of ISCC \cite{Wen2024_ISCC}. These aspects are especially critical since ISCC requires careful orchestration of sensing, communication, and computation to accomplish complex tasks. Coordination must extend beyond power, bandwidth, and computational resources to include tight synergy across heterogeneous nodes and tiers of the network, as well as seamless interaction between hardware and software modules. These interdependencies create a highly complex system that is inherently sensitive to uncertainties and disruptions.

\begin{figure*}
    \centering   
    \includegraphics[width=0.90\textwidth]{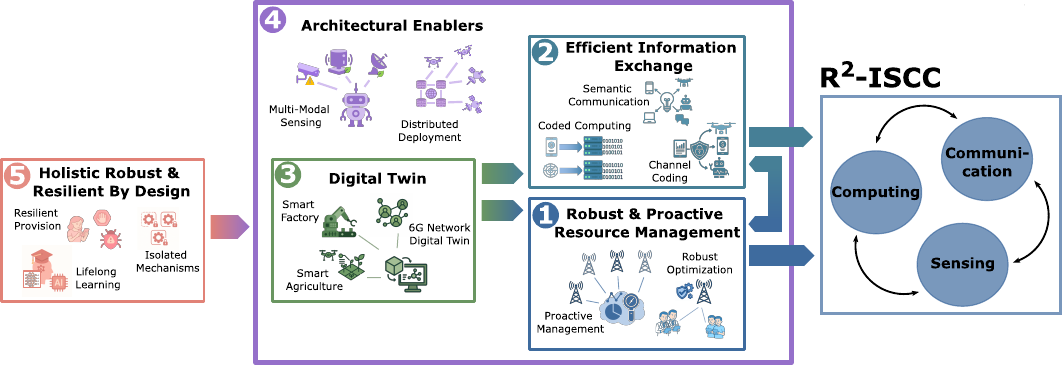}
    \caption{Key enablers of R$^2$-ISCC networks: (1) robust and proactive resource management, (2) efficient information exchange, both supported by (3) digital twins, as well as (4) architectural enablers and (5) holistic robust-and-resilient-by-design approaches.}
    \vspace{-10pt}
    \label{fig_network}
\end{figure*}

To address these issues, this article advocates the development of robust and resilient ISCC (R$^2$-ISCC) networks and discusses key design principles to achieve them.
We first introduce the fundamental frameworks for ISCC resource management and provide the related critical concepts and metrics for characterizing robustness and resilience in these networks. Next, we present five interconnected and complementary enablers of R$^2$-ISCC, namely real-time resource management and  efficient information exchange, both supported by digital twins, as well as  architectural enablers and holistic robust-and-resilient-by-design approaches, cf. Fig.~\ref{fig_network}. A case study on distributed sensing systems is presented to exemplify the design of R$^2$-ISCC networks with respect to both robustness and resilience.
Finally, we highlight key research directions and open problems for advancing R$^2$-ISCC 6G networks.

\section{Overview of R$^2$-ISCC Systems}

This section first presents a common framework for resource management in ISCC systems and their response to uncertainties and disruptions. We then discuss how robustness and resilience are measured in R$^2$-ISCC systems.

\subsection{Common Framework for R$^2$-ISCC}

\textbf{Joint design in R$^2$-ISCC:} In ISCC networks, sensing, communication, and computing share inherently limited resources such as energy, bandwidth, time, and computational cycles, often competing with one another. Joint design and resource management (RM) are therefore fundamental to coordinate these resources across devices, edge nodes, and the cloud. Treating provisioning as an integrated problem, rather than managing each domain separately, improves overall efficiency and supports diverse application tasks. This challenge is amplified in R$^2$-ISCC networks, where uncertainty requires over-provisioning and disruptions further restrict available resources. Joint design can also be task-specific \cite{Wen2024Task}, requiring adaptive and goal-driven resource allocation.
 
While integrating sensing, communication, and computing greatly expands network capabilities, it also increases operational complexity and heightens vulnerability, especially in critical applications that demand precise and timely responses. Uncertainties may stem from wireless channel randomness (e.g., fading, interference, noise), sensing and estimation errors, network dynamics, or stragglers in distributed computing. In addition, internal disruptions (e.g., unexpected failures, outages) and external attacks (e.g., jamming, physical strikes) can further compromise functionality and service availability. Therefore, there is an urgent need to design R$^2$-ISCC networks that remain robust and resilient under such uncertainties and disruptions. Fig.~1 highlights the key enablers of R$^2$-ISCC networks, which are discussed in detail throughout this article.

\begin{figure}
    \centering
    \includegraphics[width=1\linewidth]{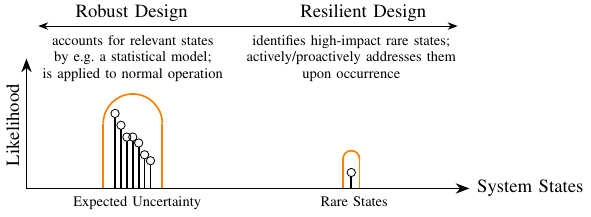}
%    \resizebox{0.95\linewidth}{!}{%
 %       \input{figure/ISCC_RR_v1.tex}
%    }
    \caption{Illustration of the continuous spectrum from robust to resilient designs, based on the rareness of disruption events and the associated system states.}
    \label{fig:enter-label}
\end{figure}

\textbf{A continuous spectrum from robustness to resilience:} Robust schemes are designed to handle uncertainties, typically using statistical models or worst-case bounds. However, they are inefficient for rare and high-impact events that require a shift in operational mode. Resilient schemes, by contrast, focus on responding to such disruptions when they occur. As illustrated in Fig.~\ref{fig:enter-label}, the transition from robust to resilient design is not sharp as event rarity and impact lie on a continuum and are partly shaped by the designer's perspective.  

To accurately characterize the robustness and resilience of the network based on its
state, the network phases illustrated in Fig. \ref{fig_resilience_phase} are considered. Specifically, the time interval between $t_0$ and $t_1$ represents a normal operating phase, during which routine uncertainties (such as channel fading, expected network errors, system imperfections, and computational stragglers) may occur.  These uncertainties can affect system performance, though not severely enough to disrupt operations.
In this phase, the ISCC network must account for the impact of such uncertainties and minimize possible performance degradation by the means of robust designs and optimization techniques. 

When the network begins to experience unexpected failures/disruptions that can drastically change its state, it enters the phase between
$t_1$ and $t_4$. Basically, the onset of $t_1$, the network performance could quickly degrade
due to these disruptions, whereby without any recovery mechanisms or intervention, the network may continue to operate at a severely diminished performance level.
However, if resilient strategies are in place, the network can begin to recover either fully to its normal state or at least improve from the degraded condition as illustrated during the period from $t_3$ and $t_4$. 
It is important to note that resilience in network design encompasses more than just recovery. It also includes mechanisms that resist disruptions proactively, limit the extent of performance degradation during disruptions, and accelerate the recovery process \cite{KhaloopourResilience6G2024}. These capabilities collectively contribute to the overall resilience of the ISCC network.

\subsection{Evaluation of Robustness and Resilience}

To evaluate the robustness and resilience of an ISCC network, several quantitative metrics are commonly used. For robustness, key indicators include outage probability and service reliability under uncertain parameters. For instance, when latency is used as a QoS metric, the outage probability can be defined as the likelihood that a task fails to meet its latency requirement given that certain conditions (such as the mean user traffic rate) are only known with limited accuracy. Other performance metrics can also be applied to assess specific aspects of the ISCC network. For example, packet and bit error rates are useful for evaluating communication performance, while detection and false alarm rates serve as indicators of sensing performance.

\begin{figure}
    \centering 
    \includegraphics[width=0.95\linewidth]{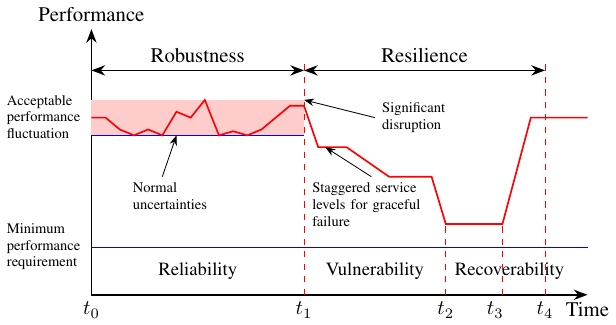}
%    \resizebox{0.95\linewidth}{!}{%
%        \input{figure/ISCC_RR_Spectrum}
%    }
    \caption{Illustration of different phases in an ISCC network with regard to robustness and resilience. This figure underlines that robustness and resilience are intertwined and characterize the system's capability to withstand a spectrum of potential uncertainties and disruptions.} 
    \label{fig_resilience_phase}
  \end{figure}

For resilience, additional compound metrics are required to capture the inherent complexity of ISCC networks and their transitions through multiple operational phases during a disruption, as illustrated in Fig.~\ref{fig_resilience_phase}.

In Phase 1, where performance begins to degrade, resilience can be evaluated based on the system's resistance capability -- its ability to maintain performance despite the onset of a disruption. One useful metric in this phase is the ratio between the time-averaged network performance under failure and that under normal conditions. Another useful metric is the worst-case performance during the disruption and service level fluctuations, which reflect how severely and variably the system degrades.

In Phase 2, where the network operates in a degraded state, the duration between 
$t_2$ and $t_3$ becomes important. This interval reflects the failure detection and response initiation time, a key indicator of responsiveness. Combined with Phase~3 (from $t_3$ and $t_4$), the total time from failure detection to system recovery ($t_2$ to $t_4$) defines the time-to-recover, a critical metric for measuring the overall resilience. Performance degradation during this interval can be further assessed using the delay in restoring system performance.

In some cases, full recovery to pre-failure performance levels may not be possible. Therefore, the post-recovery performance gap (the difference between the recovered and original performance levels) is also an important measure. Finally, over longer time spans, the mean time to failure (MTTF) should be considered to evaluate the long-term resilience of the network.

\section{Approaches for Realizing R$^2$-ISCC Systems}

This section outlines key principles for realizing R$^2$-ISCC systems, followed by concrete enabling strategies.

\label{Sec_approach}

\subsection{General Principles and Concepts}

\begin{figure}
    \centering  
     \includegraphics[width=1\linewidth]{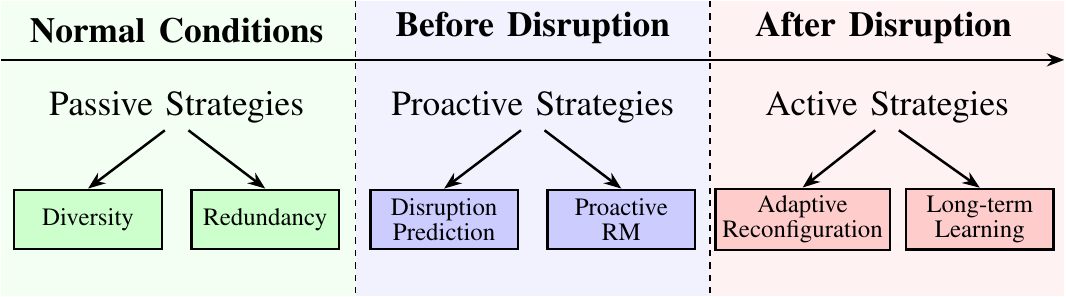}
  % \resizebox{0.95\linewidth}{!}{%
   %    \input{figure/ISCC_classes}
  % }
    \caption{Three classes of concepts for realizing R$^2$-ISCC systems. Concrete examples of these strategies are presented in Subsections III-B to III-F. }
    \label{fig_resilience_enablers}
  \end{figure}

In the following, we introduce three classes of key principles that enable R$^2$-ISCC, as shown in Fig.~\ref{fig_resilience_enablers}.

\textbf{Passive strategies:} Diversity and redundancy are fundamental from a resource management perspective. Diversity ensures multiple independent options for sensing, communication, or computation, enabling fallback when components fail. Redundancy provisions extra resources to absorb unexpected events and recover from failures. Robustness thus requires moving beyond optimization based on the average performance to also shape the tail of performance distributions and withstand worst-case disruptions.

\textbf{Proactive strategies:} Unlike passive measures, proactive strategies anticipate disruptions before they occur. By predicting uncertainties or failures, the network can take preemptive actions to avoid degradation, which is an increasingly feasible approach with integrated sensing. For example, if object tracking detects a potential blockage of a link, the transmitter can proactively adjust beamforming to maintain connectivity.

\textbf{Active strategies:} When failures do occur, protection and recovery mechanisms are essential. Resilience depends on adaptive reconfiguration, such as reassigning sensing tasks, rerouting links, or relocating computation. Limited resources further require prioritizing critical tasks and designing highly reconfigurable architectures. Over time, learning from disruptions allows the system to refine preemption, mitigation, and recovery strategies, yielding a long-term resilience.

Based on the aforementioned general concepts, the following discussion explores key enablers for realizing them within the context of ISCC systems. 

%A summary of these techniques is provided in  Table \ref{tab:summary}.

\subsection{Enabler 1 -- Robust and Proactive Resource Management}

\textbf{Robust RM:} To enhance robustness in ISCC networks, a widely adopted approach is robust resource optimization, which explicitly accounts for various network uncertainties, including target variability, channel fluctuations, computing workloads, and resource availability. The core principle is to introduce redundancy and diversity, provisioning excess resources to safeguard against worst-case scenarios. For instance, multiple base stations (BSs), edge servers, or sensors may be assigned to the same task, Thus, these tasks can be dynamically redirected to alternative resource providers with more favorable conditions. At the same time, ISCC subsystems and application tasks compete for limited resources, creating fundamental trade-offs in resource allocation. While this ensures task completion under adverse conditions, it often results in inefficient resource use, as excessive provisioning is rarely needed in typical cases. To address this, advanced robust optimization methods have been developed that focus on the trade-off between average performance and outage probability. Overall, robust RM is particularly suited to handling disruptions that commonly arise during network operation.

\textbf{Proactive RM:} Even when achieving a good efficiency–reliability trade-off, robust approaches often over-provision resources because they lack accurate knowledge of future conditions. To ensure reliability, they must prepare for a wide range of scenarios, many of which are not likely to occur. By contrast, if uncertain situations can be predicted more accurately, proactive and targeted resource management becomes possible, which allows the network to focus only on the most likely uncertainty region and greatly improving resource utilization.
Furthermore, traditional robust designs also operate within fixed tolerance levels for network fluctuations. When disruptions exceed these thresholds, failure can occur. ISCC networks, however, inherently combine sensing and computing capabilities, enabling prediction-based resource allocation to anticipate and prepare for severe events. For example, the network could preemptively activate backup servers or reserve additional capacity when a sudden disruption in computing or communication subsystems is predicted. Hence, proactive RM is able to cope with rare but predictable disruptions.

\subsection{Enabler 2 -- Efficient Information Exchange}

By allocating redundancy and diversity resources, the robustness and resilience of ISCC networks can be significantly enhanced. However, still efficiently implementing these features is crucial, ideally achieving an optimal trade-off between average performance and reliability. Below, we discuss two techniques to achieve an efficient information exchange:

\textbf{Coded Sensing-Communication-Computing:}  Channel coding is well-known for mitigating adverse communication conditions, while coded computing addresses the straggler problem often encountered in distributed computations \cite{Ng2021CodeComput}. Beyond applying coding techniques independently to communication and computation, the integrated nature of ISCC systems calls for joint coding approaches that account for dependencies across sensing, communication, and computation. Such holistic designs can further enhance overall performance and reliability despite potential disturbances affecting individual sub-systems.
At the network level, coding can also support self-repair by recovering data and functions when nodes fail. This approach is particularly effective in enhancing reliability for systems subject to independent fading states, device failures, or user misbehavior.

\textbf{Semantic and Task-Oriented Communications:} ISCC networks such as distributed surveillance, Wi-Fi sensing, and radar systems generate large volumes of raw data. With limited communication and computing resources (especially under disruptions in R$^2$-ISCC), it is crucial to exchange only task-relevant information. Semantic or goal-oriented communication meets this need by transmitting context or features directly related to the sensing, communication, or computation task, rather than raw data \cite{chaccour2024less}.

\subsection{Enabler 3 -- Digital Twin Assisted R$^2$-ISCC Networks}

\textbf{DT for R$^2$-ISCC:} A digital twin (DT) is a virtual replica of the physical environment that enables real-time simulation and analysis of system behavior \cite{Barricelli2019DT}. In R$^2$-ISCC networks, DTs can strengthen robustness and resilience by predicting network states, anticipating disruptions, and guiding adaptive responses across sensing, communication, and computation domains. For example, if a DT forecasts a power outage at a base station, the network can proactively offload communication and computation tasks to neighboring sites to sustain a minimum service. Similarly, a DT can incorporate historical traffic and current service demand to guide resource allocation (Enabler 1) or information exchange (Enabler 2) when sensing, computing, or communication subsystems fail. By combining predictive capability with context-aware decision support, DTs enhance both the anticipation and reaction mechanisms in R$^2$-ISCC.

\textbf{Robust and resilient DT:} 
Beyond enabling R$^2$-ISCC, DT itself must be robust and resilient. In fact, DTs represent a key application of ISCC networks, as they depend on sensor data collected through communication links and processed to maintain an accurate digital representation. This dependency raises key design questions about resource placement/allocation (Enabler 1) and interaction (Enabler 2) with the network. For example, a purely cloud-based DT may suffer from latency and represent a single point of failure, whereas distributed deployment at the edge brings the DT closer to users and network elements. Edge-based DTs, however, face the challenge of limited processing capacity for computationally intensive simulation, prediction, and real-time synchronization. To address this, efficient and lightweight DT models are needed, potentially combined with edge–cloud collaboration, to enable scalable, low-latency, and resilient deployment.

\subsection{Enabler 4 -- Distributed Multi-Tier Multi-Modal Network Architectures}

While Enablers 1–3 assume a fixed architecture, robustness and resilience can be further improved by architectural choices discussed below.

\textbf{Distributed sensing:} Distributed network architectures inherently enhance robustness and resilience. By spreading sensing, communication, and computation functions across multiple nodes, the failure or degradation of any single node can be compensated by others. The decentralized structure also helps localize disruptions and prevents them from cascading across the network. This makes distributed architectures particularly effective against local events such as moving blockages or physical outages.

\textbf{Multi-modal sensing:} In addition to spatial distribution, robustness and resilience can also be improved through multi-modal sensing, where different sensing modalities (e.g., camera, radar, lidar) operate in parallel \cite{Cheng2024Multi_modalSC}. Since environmental conditions typically affect modalities differently, i.e., sun glare may impair cameras but not radar while radar struggles with fine vertical motion that cameras can capture, combining complementary sensors reduces the risk of worst-case failures.  This approach is particularly effective against disruptions that degrade a single sensing modality.

\textbf{Multi-tier sensing network:} Extending the above concepts, the adoption of multi-tier and multi-layer network architectures introduces architectural redundancy and diversity, that further strengthen network robustness and resilience. or example, in dense small-cell deployments, users can connect to multiple BSs to maintain service even if one fails. With the evolution of wireless networks, services can also be supported by heterogeneous platforms such as terrestrial BSs, UAVs, and LEO satellites. Each offers distinct strengths in connectivity, sensing, and computing; when integrated through an ISCC framework, they complement one another to maintain functionality under diverse conditions. This approach is particularly valuable for large-scale disruptions such as widespread outages or natural disasters  \cite{Qiu20198UAV}.

\subsection{Enabler 5 – Holistic Approach to R$^2$BD-ISCC Networks}

A network is referred to as robust and resilient-by-design (R$^2$BD) when, already during the design phase, robustness and resilience bottlenecks (i.e., the most vulnerable components of the network) are identified, and targeted provisions (such as redundancy, adaptive coding, or reconfigurable modules) are systematically embedded into the architecture \cite{KhaloopourResilience6G2024}. In R$^2$-ISCC, sensing, communication, and computing jointly determine service quality (e.g., sensing accuracy), but differ in their vulnerability to failures. Given the cost and complexity of resilience mechanisms, a key challenge in R$^2$BD-ISCC is to identify the most critical bottleneck and prioritize both offline resource provisioning and online resource management accordingly.
To mitigate cascading failures, isolation of sensing, communication, and computing resources is essential. Furthermore, an R$^2$BD-ISCC network must support dynamic reconfiguration and the ability to shift operational modes based on the availability of functioning resources. For example, sensors may transmit high-quality raw data when high-capacity links (e.g., cellular) are available, but fall back to reduced fidelity, lower transmission rates, or locally processed results over low-capacity links (e.g., LoRa). Likewise, edge inference algorithms must adapt to variations in data quality and even operate with missing sensor inputs. An R$^2$BD-ISCC foresees such contingency operational modes during the design phase, and develops reconfiguration strategies to enable seamless transitions among them.

Fig. 1 illustrates the synergies among the proposed enablers of R$^2$-ISCC networks. In a nutshell, Enabler 1 manages the distribution of limited resources in R$^2$-ISCC networks, Enabler 2 maximizes their use through efficient information exchange, and Enabler 3 provides contextual knowledge to support both. While these three operate within a given architecture, Enabler 4 introduces architectural choices that further enhance robustness and resilience. Finally, Enabler 5 ensures that provisioning decisions are made holistically at the design stage, which directly shapes  Enablers 1–4.

\section{Case Study and Open Problems}

In this section, an illustrative case for an R$^2$-ISCC network is first presented, which is then followed by various open problems.

\subsection{Case Study}

\begin{figure}
    \centering
    \includegraphics[width=0.95\linewidth]{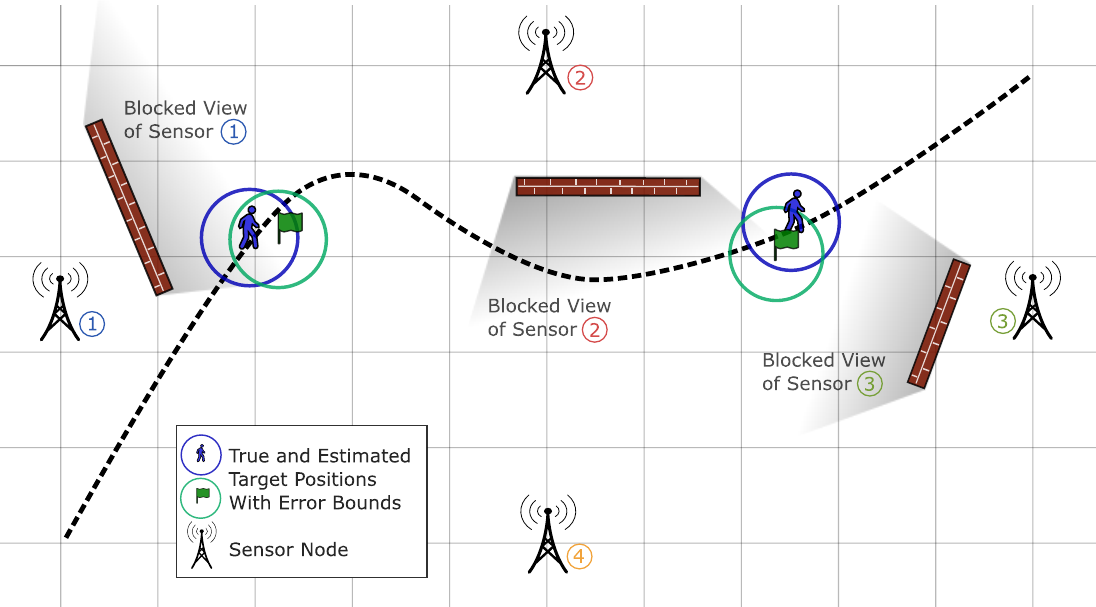}
    \caption{Illustration of a distributed sensor setup for cooperative target tracking. The true target location  is uncertain (within blue circles); the estimated location (flag) includes an uncertainty region (green circles). Moreover, sensor views may be obstructed illustrated by walls.}
    \label{fig:SimSet}
\end{figure}

As discussed in Section III, realizing an R$^2$-ISCC requires a design that is both robust to parameter uncertainties and resilient to disruptions. To illustrate the interplay between robustness and resilience, we consider sensor selection and power adaptation in a distributed radar network \cite{eckrich2024fronthaul}. The optimal radar is chosen based on the target’s \textit{estimated} location. To meet a required sensing SNR~$\gamma_{\mathrm{req}}$, sensor~$i$ transmits with power $P_i(n) \propto \hat{d}_i^{4}(n)$ under a LoS mono-static radar model, where $\hat{d}_i(n)$ is the \textit{estimated} distance to the target at time~$n$. For generality, we do not adopt a specific localization or tracking algorithm. Instead, their effects are abstracted through two factors: (i) random location estimation errors of up to $\epsilon_{\max} = 1$~m, and (ii) loss of tracking due to fixed blockage regions, as illustrated in Fig.~\ref{fig:SimSet}.

\begin{figure}
    \centering
    \includegraphics[width=1\linewidth]{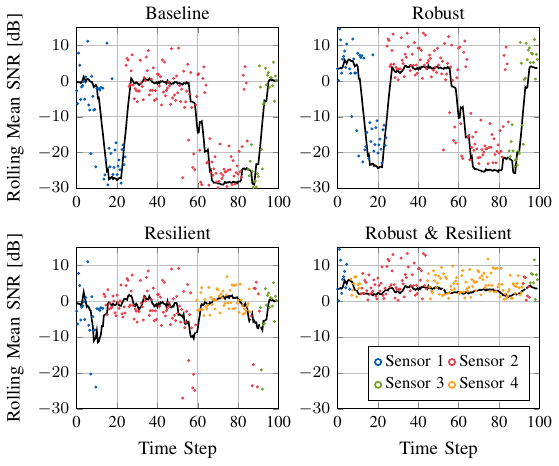}
   % \resizebox{1\linewidth}{!}{%
   %     \input{figure/simulation_results_review}
   % }
    \caption{Time evolution of the mean SNR (dB relative to target SNR) for sensors S1 to S4 under standard, robust, resilient, and robust \& resilient schemes. The window length of the rolling mean is 10 epochs. The colored circles indicate the instantaneous SNR and which sensor has been selected.}
    \label{fig:SimRes}
\end{figure}

Fig.~\ref{fig:SimRes}  shows the sensing SNR (normalized to $\gamma_{\mathrm{req}}$) over 200 epochs along the trajectory in Fig.~\ref{fig:SimRes} , comparing four sensor-selection and power-adaptation strategies discussed below.

\paragraph{Baseline}
The baseline scheme selects the sensor with the smallest estimated distance and sets transmit power based on this point estimate, ignoring both location uncertainty and blockages. As observed from Fig.~\ref{fig:SimRes}, tracking fails when the target enters a shadowed zone. Even in free space, the SNR requirement is not consistently met due to underestimated power caused by location errors.

\paragraph{Robust}
The robust design accounts for worst-case location error in both sensor selection and power adaptation. This ensures that the sensing requirement (i.e., above $0$~dB) is consistently satisfied in unobstructed segments, though at the expense of higher average transmit power. However, this scheme still suffers from sudden deterministic SNR drops caused by neglected blockages.

\paragraph{Resilient}
The resilient design proactively responds to disruptions by selecting sensors with an unobstructed view, thereby avoiding severe SNR drops due to blockages. Nonetheless, random SNR drops occur when the true location lies in a shadowed region but the estimated location does not, highlighting a lack of robustness to location errors.

\paragraph{Robust \& Resilient}
The combined robust-and-resilient scheme simultaneously accounts for location uncertainty and blockages. As a result, the sensing SNR remains reliably above the threshold throughout the trajectory.

This example illustrates how location errors and blockages jointly affect ISCC performance, and how R$^2$-ISCC strategies mitigate them in a simplified setting. Real-world ISCC systems are far more complex, involving various uncertain parameters as well as both known and unknown disruptions. This highlights the need for further research on R$^2$-ISCC, with key directions reviewed next.

% The relative distances is to scale; error circles are scaled by a factor of two for readability.

\subsection{Open Problems}

To complete this paper, some critical open problems for R$^2$-ISCC are discussed below.

\textbf{Efficient and scalable resource management for ISCC:} Although widely used in conventional networks, diversity and redundancy schemes are typically designed for individual purposes, without jointly addressing sensing, computing, and communication. In ISCC networks, new strategies that can efficiently integrate diversity and redundancy across all three components are needed, particularly for tasks that exhibit intrinsic dependencies among these functions. Moreover, the joint optimization of sensing, computing, and communication resources results in inherently high-dimensional and complex optimization problems. Hence, the development of efficient, scalable solution methods is critical to enable practical deployment and operation of robust and resilient ISCC systems.

\textbf{Task-oriented and criticality-aware ISCC:} ISCC networks must support heterogeneous tasks with varying requirements and urgency. To improve robustness and resilience, proactive resource management must account for these task-level differences, especially during disruptions when resources are scarce and prioritization is vital. Network slicing offers a promising solution, but conventional methods focus mainly on communication. In ISCC, holistic slicing approaches that jointly manage sensing, communication, and computation resources are needed.

\textbf{Joint source–channel–computing network coding schemes:} To support network-level resilience, coding must extend beyond individual node types and retain the network functionality even when some sensing, communication, or computation nodes become unresponsive or fail. For example, coded computing techniques should be adapted to the ISCC setting to account for failures such as missing sensor inputs. Similarly, network coding, known for its resilience to link failures, eavesdropping, and jamming, offers a promising foundation for designing ISCC architectures capable of withstanding diverse and unpredictable network-level disruptions.

\textbf{Synergies of multi-tier multi-modal ISCC networks:} Distributed multi-tier, multi-modal ISCC networks enhance robustness and resilience through complementary fault-tolerance features. However, research in this area is still at an early stage, and require new architectures and design methodologies to integrate heterogeneous devices across locations. Effective fusion techniques are also needed to align feature spaces across sensing modalities for cooperative perception. From a multi-task perspective, designing AI and perception tasks to share a unified feature space would allow a single joint data representation to serve multiple tasks, which could greatly reduce communication overhead.

\textbf{Resilience bottleneck in R$^2$BD-ISCC} networks: The R$^2$BD principle emphasizes embedding resilience capabilities during the design phase. In ISCC networks, resilience provisions across sensing, communication, and computing components are inherently heterogeneous and contribute differently to the end-to-end service resilience. Efficient provisioning requires identifying resilience bottlenecks and measuring the end-to-end resilience based on the performance metrics discussed in Section II-B, so that investments can be directed to the components that yield the greatest improvement in overall resilience. While the impact of resilience provisioning in individual subsystems has been studied, a holistic, joint approach to R$^2$BD-ISCC networks remains largely unexplored.

% This challenge is further amplified in task-oriented designs, where optimal provisioning trade-offs depend heavily on specific task objectives.

\textbf{Self-organizing ISCC networks:} A critical aspect of the R$^2$BD principle is reconfigurability based on available resilience measures. However, to hugely benefit from the network reconfigurability, corresponding strategies that support effective task redistribution through resource reallocation and network reconfiguration are required. Furthermore, while existing self-organizing networks (SONs) in communication systems have partially addressed this need, they do not account for the specific requirement of ISCC networks, which involve far more complex and tightly coupled sensing, computing, and communication operations \cite{Chaoub2023SON}.

\section{Conclusions}
ISCC networks are essential for enabling intelligent perception in domains such as networked robotics, smart healthcare, Industry 5.0, and autonomous transportation. However, the interdependencies among sensing, computing, and communication components make them inherently complex and more vulnerable to uncertainties and disruptions. To address this, robustness and resilience must be embedded into their design. This article proposes an R$^2$-ISCC framework, and outlining key principles and strategies across resource management, coding, DTs, architecture,  and network-level design. A simulated case study highlights the importance and interplay of robustness and resilience. Key research problems are identified to guide future work. We hope this study inspires continued exploration and innovation toward R$^2$-ISCC networks.

%\appendices

%紙本用alpha,abbrvnat   電子本用alphaurl
%\bibliographystyle{abbrvnat}
\balance
\bibliographystyle{IEEEtran-et-al}
\bibliography{Bib_ISCC_Mag_2025_04}

\vspace{0.3cm}
\noindent 
\textbf{Ming-Chun Lee} (Member, IEEE) is an Associate Professor in the Institute of Communications Engineering at National Yang Ming Chiao Tung University (NYCU), Taiwan. His research interests include wireless sensing and communications of systems and networks.

\vspace{0.3cm}
\noindent 
\textbf{Christian Eckrich} (Student Member, IEEE) is a Ph.D. student at the Resilient Communication Systems (RCS) Lab, Technical University of Darmstadt (TUDa), Germany. His research interests include distributed radar systems, through-wall sensing, and vital sign detection.

\vspace{0.3cm}
\noindent 
\textbf{Vahid Jamali} (Senior Member, IEEE) is the Head of the RCS Lab at TUDa. His research interests include wireless communications and sensing, resilient communication systems, and bio-inspired molecular communications.

\vspace{0.3cm}
\noindent 
\textbf{Yu-Chih Huang} (Senior Member, IEEE) is a Professor in the Institute of Communications Engineering at NYCU. His research interests include information theory, coding theory, wireless communications, and statistical signal processing. 

\vspace{0.3cm}
\noindent 
\textbf{Arash Asadi} (Senior Member, IEEE) is an Assistant Professor in the Embedded Systems Group at Technical University of Delft, Netherland. His research interests include wireless communication and sensing for beyond-5G/6G networks. 

\vspace{0.3cm}
\noindent 
\textbf{Li-Chun Wang} (Fellow, IEEE) is a Chair Professor and the current dean of the College of Electrical and Computer Engineering at NYCU. His current research interests are in the areas of radio resource management and cross-layer optimization techniques for heterogeneous wireless networks, and cloud computing for mobile applications.

\end{document}